# GENERATIVE ARTIFICIAL INTELLIGENCE IN SMALL AND MEDIUM ENTERPRISES: NAVIGATING ITS PROMISES AND CHALLENGES

**Kumaran RAJARAM (corresponding author)**
Nanyang Technological University, Singapore
Nanyang Business School
Division of Leadership, Management and Organization
91, Nanyang Avenue, Gaia, ABS-05-003
Singapore 639956
rkumaran@ntu.edu.sg
+65 94553150

**Patrick Nicolas TINGUELY**
ETH Zurich, Switzerland
Department of Management, Technology, and Economics
Chair of Strategic Management and Innovation
Weinbergstrasse 56/58
8092 Zurich, Switzerland
ptinguely@ethz.ch
+41 44 632 08 19

# GENERATIVE ARTIFICIAL INTELLIGENCE IN SMALL AND MEDIUM ENTERPRISES: NAVIGATING ITS PROMISES AND CHALLENGES

**ABSTRACT**

The latest technological developments in generative artificial intelligence (GAI) offer powerful capabilities to small and medium enterprises (SMEs), as they facilitate the democratization of both scalability and creativity. Even if they have little technical expertise or financial resources, SMEs can leverage this technology to streamline work processes and unleash innovation, thereby improving their product offerings and long-term competitiveness. This paper discusses how SMEs can navigate both the promises and challenges of GAI and offers a roadmap for deploying GAI. We introduce a sailing metaphor that reveals key strategic dimensions for GAI deployment: competency of employees, effective leadership and work values, organizational culture, collaboration and cooperation, and relationships with third parties. We offer practical recommendations that serve as a useful compass for successfully deploying GAI in SMEs.

**Keywords:** Generative Artificial Intelligence; Small and Medium Enterprises; AI Management; Competitiveness; Innovation.

# THE RISE OF GENERATIVE ARTIFICIAL INTELLIGENCE

The latest advancements in generative artificial intelligence (GAI), a technology that can create content (e.g., text, image, video, code) based on patterns from large training datasets (Jebara, 2004), have attracted significant attention among practitioners and researchers alike. GAI has become popular following the release of a conversational chatbot called ChatGPT (Generative Pre-Trained Transformer) in November 2022 (OpenAI Blog, 2022), which became the fastest-growing consumer application in history, with 100 million active users within two months after its launch (Hu, 2023). In most organizations, employees at any level or function can already access and experiment with the powerful capabilities of GAI, regardless of their technical background or experience. While most of the current attention revolves around the short-term applications of such novel technology (e.g., searching or synthesizing information online more efficiently, generating creative images through prompts), GAI is poised to profoundly transform industries and organizations in the long term (Gates, 2023; Felten et al., 2023).

GAI will disrupt the way organizations operate, as it is the first technology in human history that generates its own content rather than disseminating or supporting content created by humans (Milmo, 2023). Unlike earlier AI models powered by machine learning (ML) algorithms, GAI draws from more complex inputs and generates entirely new content that is human-like, highly coherent, and creative. While discriminative AI specialized in using existing data for content categorization (Jebara, 2004), GAI relies on large language models (LLMs) to create new data by predicting the likelihood of the appearance of a particular next word in a text, conditional on the lexical context. LLMs belong to the latest generation of AI foundation models, which are pretrained on large sets of generalized and unlabeled data and can be fine-tuned to accommodate a broad range of tasks, such as conversing in natural language or generating novel content Once those models are trained, individuals can "prompt" them to direct their response to specific tasks without additional data or any task-specific training (Kojima et al., 2022).

GAI is a game changer for small and medium enterprises (SMEs), because it offers state-of-the-art capabilities that used to be reserved to the largest firms, thereby democratizing scalability and creativity. As SMEs are fundamentally agile, they can swiftly adopt GAI to close the knowledge and technological gaps with larger firms, thus leveling the playing field (Acar & Gvirtz, 2024). Representing approximately 90% of businesses and more than 50% of employment worldwide (The World Bank, 2023), SMEs play a pivotal role in contributing to the economy across industries, particularly in developing countries.

While GAI can deeply transform how SMEs operate, scholarship on AI management (e.g., Kaplan & Haenlein, 2020; Leyer & Schneider, 2021; Paschen, Pitt, & Kietzmann, 2020; Lee & Shin, 2020) falls short of providing actionable recommendations for deploying the technology. First, because the latest LLMs have only recently emerged, research has not yet examined how organizations can handle the idiosyncratic challenges involved in their deployment. LLMs typically involve more complex inputs than prior AI iterations and result in novel human-like outputs, which makes them more flexible and scalable. They mainly impact occupations that involve non-routine, highly cognitively complex tasks, by contrast to prior AI advancements, which disrupted primarily routine tasks (Felten et al., 2023). Past work has focused on AI systems that perform routine, boring tasks that workers are not willing to do (e.g., Kaplan & Haenlein, 2020). As GAI can surpass humans in performing tasks that are creative and involve sense making or decision making, it has the potential to create frictions that may compromise its social acceptability (Tinguely et al., 2023). For instance, GAI can threaten employees' professional identities (e.g., Lebovitz et al., 2022), which calls for guidance on the adoption of GAI.

Second, the practitioner-oriented literature fails to address the specific challenges of SMEs, because it focuses on AI deployment in large organizations. SMEs lack the breadth of complementary resources that facilitate GAI deployment, such as human skills (in e.g., data science) and hardware and software architecture (Bughin, 2023). While we recognize that most SMEs will merely be sophisticated users of GAI, some may want to customize existing models to better meet their business needs, which requires some effort in GAI development. Furthermore, SME employees typically wear multiple hats and need to handle the many aspects of GAI deployment with little support or guidance. SME leaders tend to oversee multiple functions of the SME, and they may be responsible for GAI's seamless integration across different departments. As teams are relatively small and hierarchies remain typically flat in SMEs, cross-functional collaboration is essential to gather relevant insights and align GAI initiatives with the business goals. Finally, SMEs typically lack established processes and infrastructure that support business relationships with external parties. This set of unexplored SME specificities in AI deployment calls for a better understanding of the unique competencies required for SME employees and their leaders.

In this paper, we present both the promises and the challenges of GAI for SMEs. We tackle the following crucial and timely research question: How can SMEs deploy GAI that is well accepted by all stakeholders and has the potential to unleash scalability and creativity? In our exploration, we develop a sailing metaphor that lays out the key strategic dimensions for GAI deployment: competency of employees, effective leadership and work values, organizational culture, collaboration and cooperation, and relationships with third parties. We articulate practical guidance that serves as a useful compass for successfully deploying GAI in SMEs. We conclude by revealing how our work contributes to scholarship on AI management.

## SMES ARE ENTERING UNCHARTED, CHOPPY WATERS

SMEs are entering uncharted waters, because GAI is an emerging technology whose features are still in flux (e.g., Bailey et al., 2022), implying that novel use cases are evolving fast with the latest advancements of LLMs. Further, the waters are choppy, because the rise of GAI is disrupting the competitive landscape of SMEs. Many will disappear if they fail to leverage GAI to improve scalability and creativity. This section explores how GAI is disrupting work within SMEs and summarizes the promises and challenges of GAI for SMEs.

### GAI Is Disrupting Work within SMEs

GAI is transforming how SMEs organize and perform work activities. Occupations affected the most by GAI are the ones performing non-routine, highly cognitively complex tasks (Felten et al., 2023), which mainly concerns highly educated, highly paid white-collar employees. GAI will need to collaborate with employees in those occupations to fully augment their capabilities (Chui, Roberts, & Yee, 2022). GAI can transform interaction labor (e.g., customer service) by mimicking human interactions in very subtle ways, yet most use cases may still require human inputs and intervention. For instance, chatbots built on LLMs can reshape how businesses engage their customers (Kietzmann & Park, 2024), address specific customer needs (e.g., product comparison and selection), and improve conversion rates (Jain & Ritwik, 2023). The latest GAI models enable versatility, scalability, and creativity in their use cases, which makes them relevant across many occupations and industries where SMEs operate. For example, in financial services, GAI can explore different investment strategies and recommend customized and unique financial models to customers. It can also help predict future performance for the SME by assessing the evolving market conditions, anticipating market sentiment, and extracting real-time insights by sieving through a huge amount of social content. In the healthcare industry, GAI models can support scalable solutions in terms of healthcare decision making through data analytics. They can also provide accurate diagnoses, personalize medical treatment plans based on individual patient characteristics, foretell potential disease outbreaks, and support the design of novel treatment methodologies. For instance, SMEs in precision medicine can leverage GAI to customize models and treatment plans using the individual physiological data and family medical history that are available in electronic forms (Kuzlu, Xiao, Catak, Gurler & Guler, 2023).

Sustaining the growth of their customer base and the internationalization of their markets is key to SMEs' survival and requires solutions that address such complex challenges. SMEs may embrace GAI, as it offers creative and interactional capabilities that have thus far been limited to large firms with deep pockets. Early evidence suggests that small businesses find GAI helpful in tasks that are particularly challenging given a lack of time or skills, such as attracting new clients, growing revenue, or creating engaging digital marketing content (GoDaddy Inc., 2023). For instance, a small Swiss chocolate confectioner can use GAI to customize its advertising material and translate it to many languages to scale up its business globally. SMEs can create customized chatbots that provide 24/7 human-like responses to clients' requests, which improves customer satisfaction and retention at low cost (Forbes, 2023; Ferraro, 2024). Hence, powerful LLMs can provide SMEs with cost-effective and time-efficient solutions to unleash creativity or deliver tailored and exceptional experiences to their customers (AI TechPark, 2023).

**GAI Promises and Challenges for SMEs**

One notable strength of GAI lies in the automatic generation of content, which can take two different forms (Felten et al., 2023). First, advances in language modelling mean that generative writing tools like ChatGPT can automatically create original content, such as market reports, with little human intervention. Second, novel developments in image generation mean that tools such as Midjourney, DALL-E, or Stable Diffusion can generate images from text descriptions (prompts) and augment the work of digital designers by generating compelling visuals for a marketing campaign.

Given those strengths, GAI is rapidly gaining traction in marketing, with the top three application domains being customization, insight generation, and content creation (Kshetri et al., 2023). In those domains, GAI can complement previous AI iterations, which have enabled a wide range of human-machine integration. For instance, AI is already being used in advertising and customer experience for generating valuable insights into target consumers and their engagement or conducting return-on-investment analysis (Campbell, Sands & Ferraro, 2020). AI has changed the way advertisers comprehend and guide consumers, for instance through data mining that delivers consumer insights on the spot (Kietzmann, Paschen & Treen, 2018). The marketing insights generated by the latest GAI applications are highly personalized and more creative than the ones produced by previous generations of AI. Hence, GAI can dramatically increase the effectiveness of consumer marketing and service (Paul et al., 2023).

Early quantitative evidence points to the benefits of employees' exposure to GAI (for example, using tools like ChatGPT), as it is associated with potential productivity gains for firms (Eisfeldt et al.,2023). In this paper, we posit that SMEs are good candidates for integrating GAI into their business operations, because GAI can enhance both scalability and creativity. Their size typically restricts them from realizing large economies of scale, as they cannot spread costs (e.g., human resources, marketing, legal, and accounting) over a large volume of products and services. GAI can greatly improve efficiency for tasks underlying those costs (e.g., creating automatic monthly reports in analytical accounting), which reduces fixed costs and hence unit production costs of products and services. Automating those tasks frees up employee time and resources, which can be reallocated to other promising areas. SMEs can also leverage GAI to generate creative ideas, prototypes, and products and services, thus improving the way they meet customers' needs. The relatively low cost of third-party GAI tools makes this technology ideal for SMEs with constrained budgets. In Table 1, we briefly describe and provide practical examples of six key GAI promises that may lead to competitive gains for SMEs: transforming roles and boosting productivity, accessing human knowledge, facilitating creativity and empowering innovation, increasing differentiation of products and services, accelerating product development, and supporting decision making.

**[Insert Table 1 about here]**

Despite its promises, GAI presents important challenges that arise from the very characteristics of ML algorithms powering GAI or from the way SMEs adopt such technology. For instance, marketing professionals may be reluctant to use GAI to design advertising visuals, because they may strongly identify with such tasks and because such tasks give them visibility and status within the SME. They may feel threatened if GAI can perform these tasks. Another important challenge is that GAI models are prone to "hallucinations," where their outputs contain "random inaccuracies or falsehoods expressed with unjustifiable confidence" (Mukherjee & Chang, 2023; Hannigan et al., 2024). For instance, a New York attorney from a small law firm relied on a GAI chatbot to extract case law citations for a personal injury case against an airline (Weiser, 2023). The chatbot provided legal cases that were fabricated and could not be found in reputable legal databases and the judge eventually fined the attorney for failing to verify his sources.

We classify the key GAI challenges for SMEs into six categories: adoption hurdles, accuracy of predictions, ethical issues, reputational and legal risks, reliance on third parties, and financial burden. In Table 2, we present the challenges of GAI for SMEs and provide practical examples for each of them.

[Insert Table 2 about here]

## NAVIGATING GAI DEPLOYMENT IN SMES: THE SAILING METAPHOR

We develop the sailing metaphor to frame our discussion and illustrate our insights. In this metaphor, companies are ships that navigate the waters of a changing competitive landscape. Fast technological progress typically comes with profound regulatory and societal changes, which creates discontinuities in the competitive landscape and presents navigation challenges for the most vulnerable ships (i.e., SMEs). Given their limited capabilities, SMEs are particularly ill-equipped to cope with strong waves. They can only navigate those uncharted, choppy waters visually. GAI creates an unprecedented disruption (i.e., a heavy storm) that destabilizes all ships instantly, requiring them to constantly adapt their sailing approach to stay afloat. GAI pushes SMEs to radically change how things have been done in the past. SMEs need to realize that GAI is a technological disruption that will have vital repercussions for their business.

In sailing, four key aspects broadly influence a ship's ability to get from Point A to Point B smoothly and safely: the quality of the crew, the quality of the ship captains, the navigation of the ship, and the sailing route. Table 3 presents the application of the sailing metaphor to SMEs that aspire to deploy GAI. We describe the strategic elements and the required competencies for GAI deployment in light of the SME specificities. In Figure 1, we leverage GAI ourselves to illustrate the sailing metaphor (which also serves as a flagship example of how GAI can be creatively adopted by management researchers).

[Insert Table 3 about here]

[Insert Figure 1 about here]

## The Quality of the Crew

The first element of the sailing metaphor is the quality of the crew, which relates to the competency of SME employees.

**Competency of Employees**

Employees need to have three key sets of competencies to cope with and resolve the important challenges of GAI deployment and the discomfort that arises, namely a learning orientation, technology savviness and curiosity, and adaptability. First, SME employees need to learn on the fly with few structured training programs. Working in a small structure, SME employees need to creatively find solutions to most issues and learn by themselves (Haryanto, Hatyono, & Sawitri, 2017). Thus, they need to exhibit a strong learning orientation, which refers to their ability to continuously unlearn and learn to work in such a new system. Second, SME employees are typically directly involved in the GAI deployment and have some degree of decision-making authority over GAI use (Gonzalez-Varona, Acebes, Poza, & Lopez-Paredes, 2020). Therefore, they will need the ability and curiosity to experiment with novel technological features and stay up to date with rapidly evolving technological developments and the latest advancements in GAI tools, use cases, and policies. This includes experimenting with GAI for novel use cases and engaging in relevant professional development workshops, conferences, industry exchanges and training programs. Technology-savvy employees must understand the cyber security implications of GAI systems, for instance by being aware of potential vulnerabilities and exemplary practices in protecting data and applications. Because SME employees have several different roles in the organization and handle many aspects of GAI deployment (Wishart, 2018), they need to demonstrate a high degree of adaptability, which is the ability to realign and meet the rapidly evolving needs arising from GAI deployment.

SME employees need to stay adaptable with an agile mindset to acquire new skills in collaborating with GAI systems, as integrating GAI in workflows disrupts job responsibilities and roles. This involves understanding how to interpret and provide human expertise for GAI outputs.

## The Quality of the Ship Captains

The next element of the sailing metaphor is the quality of the ship captains, which refers to effective leadership and work values.

**Effective Leadership**

SMEs' flexible structures and processes allow them to smoothly embrace technological change, yet their opportunities for value creation are constrained by tight resources. SME leaders need to make crucial strategic decisions on resource allocation, accounting for cost and potential (Walt, Kroon & Fourie, 2015). Therefore, SME leaders need to develop a clear vision on how to leverage the opportunity of GAI's powerful capabilities for scalability, creativity, and business growth within tight resource constraints. SME leaders typically oversee multiple functions of the SME and must ensure a seamless GAI integration across different departments and good alignment with the broader business goals (Menkhoff, Loh & Kay, 2023). A leader in the role of a change agent inspires confidence and empowers others to take ownership in the transition phase, raising their spirits and reassuring them of the direction, which incites employees to experiment and expand their present views and meaning-making processes, enhancing their capability to think in a more systemic and interdependent way. Leaders as change agents are effective ambassadors that empower employees to appreciate the benefits of GAI for scalability and creativity. They can encourage an innovative mindset to view and embrace GAI as beneficial to all stakeholders. Furthermore, SME leaders must quickly respond to evolving market conditions by taking advantage of the flexible structures and processes of the SME (Walter, 2021). They need to be agile in the GAI deployment phase by quickly addressing GAI's disruptions in the workflow with focused and high-quality solutions.

**Work Values**

SMEs often juggle contrasting demands from stakeholders, which arise from varying markets, regulations, and ethical standards and result in enduring and deep-rooted tensions (Askeland et al., 2020). Work values of SME leaders can help them navigate complex and unprecedented circumstances with consistency. As GAI deployment leads to changes in workflow and redeployment of resources, leaders' work values help them make informed decisions. SME leaders need to shape the mindset and behavioral norms of their employees to help them cope with the disruptions caused by GAI deployment that affect their work routines and roles.

SME leaders must have the foresight to understand customers' evolving specific needs and expectations so that they can leverage GAI's powerful capabilities to enhance the firm's business growth by optimizing the constrained resources available. For instance, Textio is an SME that relies on GAI to guide recruiting teams in developing specific profiles for candidate outreach and to improve the quality of job postings to speed up the recruitment process and attract quality job applicants. Textio's GAI platform examines language patterns and trends and predicts the type of job descriptions that will attract a quality talent pool in a particular industry. This customized approach for job postings assists firms to attract the right job-fit talent quickly and more accurately. It partners with the collaborating firm to offer tailor-made, bias-free, actionable, and equitable performance reviews for their employees. Textio also helps managers to deal with systemic issues rooted in unconscious bias (e.g., exclusion and inequity at work) by analyzing data insights and offering customized solutions (Textio, 2024). For these functionalities to be effectively incorporated and operated, leaders need insight and the ability to resonate with the customers' rapidly changing expectations. Furthermore, SME leaders must be agile in implementing GAI in the workflow process by working around their limited resources and talent capabilities to be able to provide customer-oriented solutions. For instance, Tastewise (TasteGPT) uses GAI to curate tailor-made menus for restaurants. Leveraging food trends and consumer data analysis, Tastewise deploys GAI algorithms to recommend menu items customized to the tastes and preferences of each of its clients. This highly personalized strategy becomes an enabler for restaurants to differentiate themselves in a highly competitive market (Tastewise, 2024). The successful execution of such a strategy requires an SME leader to be equipped with specific work values such as agility and responsiveness.

## The Navigation of the Ship

The third element of the sailing metaphor is the navigation of the ship, which includes the organizational culture, collaboration, and cooperation across and within hierarchical levels.

**Organizational Culture**

Organizational culture is a system of beliefs, values, and behavioral patterns that subconsciously influences employees in their decision making (Ortega-Parra & Sastre-Castillo, 2013). It shapes how employees interact among themselves and with other stakeholders (Simoneaux & Stroud, 2014) and can play a vital role in how smoothly SMEs incorporate GAI into their workflow. We discuss two types of organizational culture and how they influence technological adoption. First, a market culture is inclined towards addressing contentions, market evolution, and attainment via the firm's goal deliverables (Deshpandé, Farley and Webster, 1993). For instance, GreenTech Innovations is a dynamic SME operating in the renewable energy sector. The firm has effectively cultivated a market culture among its employees, driving creativity and excellence by aligning its strategic direction towards innovation and sustainability. The market culture is nurtured and ingrained through varying approaches. For instance, the SME involves their sales team in contests every month. The sales team members compete in attaining the highest sales figures, with the top achievers rewarded with bonuses and offered public recognition at company-wide events. Furthermore, to address evolving changes, GreenTech Innovations organizes regular innovation challenges where cross-functional teams compete to develop novel, eco-inclined technologies and innovations. For instance, they organized an internal competition to draw feasible designs for a more efficient solar panel, with the winning team's design adopted for the next product line (Greentech Innovators, 2024). This example illustrates that a market culture can be a driving force that makes SMEs embrace technological adoption.

Second, an adhocracy culture is innovative and adaptable (Veiseh et al., 2014). In contrast to bureaucratic cultures, it fosters individual initiative, decentralized decision making and leadership. Adhocratic SMEs across many industries, such as aerospace, gaming or automotive, can adopt GAI to foster innovation, such as facilitating the customized design of parts. Employees evolving in such an organizational culture are more receptive to technological advancement, as they have a strong change tolerance mindset, which enables them to display positivity with little resistance in approaching uncertainty and disruptions. For instance, Valve Corporation is an SME in the video game development and digital distribution field that has adopted an adhocracy culture through its unique approach to organizational structure and decision making. The firm is well known for doing away with a traditional hierarchy and operates with a flat organizational structure without managers. Employees are encouraged to be involved in decision-making processes by taking on projects they are ardent about. Collective consensus is Valve's approach to the decision-making process. A strong emphasis is placed on open communication and feedback where individuals and teams work together to make decisions. Valve creates and fosters an innovative environment that empowers employees to experiment with novel ideas and pursue projects in their interests, which significantly fosters technological adoption (Valve Corporation, 2024).

**Collaboration**

As the organizational structure of SMEs is rather lean, with small teams and flat hierarchies, SME employees and leaders need to collaborate closely and effectively in GAI deployment (Major, 2023; Mohiuddin, 2017; Karadakal, Goud & Thomas, 2015). Collaboration entails working together with a common or shared goal and involves joint efforts and contribution from all parties involved. Pursuing collective goals is crucial to identify and cope with the challenges arising from GAI adoption. A relational commitment and/or quality engagement is required to work together with varying cross-departmental colleagues within the organization and incorporate GAI work processes. Effective collaboration requires employees' conflict-management, problem-solving, and negotiation skills. Employees need to deal with and address issues arising from GAI disruptions with tact yet objectivity. They need to find common ground and reach a win-win situation addressing everyone's needs. In simple terms, SMEs enable collaboration through meaningful information sharing, distributed resources, and defined shared responsibilities to collectively analyze, plan and implement the activities required to attain common goals.

**Cooperation**

Cooperation entails working towards a common goal while the parties involved maintain their individual focus, retaining a degree of autonomy and independence. It includes managing interdependencies between individuals and teams, for instance by addressing the responsibilities assigned in a written contract and the intangible contributions beyond it. For instance, in material science GAI is having a significant impact on the medical, energy, electronics, automotive, defense, and aerospace industries by creating new materials with explicit physical properties. A unique process called inverse design defines the required properties of materials and unveils materials that potentially contain the essential properties rather than relying on chance. In this example, cooperation is crucial, as employees need to connect with interdepartmental colleagues, leveraging their unique expertise towards a common purpose. Good cooperation requires that employees demonstrate alignment and team building skills, bringing everyone together to see the common goal and holistic benefit of deploying GAI in the SME.

## The Sailing Route

The next element of the sailing metaphor is the sailing route, which refers to the SME's chosen itinerary for collaborating with third parties.

### Relationships with Third Parties

Resource-constrained SMEs can collaborate with third parties to gain access to advanced and complex GAI development tools, hardware, and software infrastructure (Audretsch & Belitski, 2020). They can heavily rely on multiple GAI solution providers for computing services (hardware and software), data services (creation, collection, preparation), foundation modeling services (closed-source via an API such as GPT-4 and open-source foundation models such as CLIP), and application development (software, devices, or applications for end users) (Höppner & Streatfeild, 2023). SMEs can also partner with research institutions to access cutting-edge research, talent, and expertise and stay updated with the latest technological advancements. They can collaborate with experts or organizations who have the domain knowledge and experience in their specific field, which can assist them in customizing GAI solutions to industry-specific challenges.

Alternatively, SMEs can build on the competitive advantage of business partners that have adopted competence-enhancing innovation through AI (Paschen, Pitt, & Kietzmann, 2020). They can enter strategic alliances with other companies in the GAI space to pool resources, share knowledge, and work on GAI projects that are mutually beneficial. An important boundary condition is that SMEs are restricted in which business partners they can collaborate with because of their size and capabilities. Hence, they could for instance pool resources with other SMEs to personalize chatbots addressing clients' queries that are specific to their industry.

A successful relationship with third parties requires competencies in project management, risk management, and communication. First, SMEs typically lack established processes and infrastructure that support collaboration with external parties (Meister, 2006). Hence, they need to define clear goals and milestones and monitor timelines to make sure that GAI deployment with external parties stays on track. Second, due to their entrepreneurial mindset and minimum bureaucracy, many SMEs have limited resources and expertise in risk management, contracting and compliance for developing novel technological solutions with third parties (Bluebox Content Team, 2024). Therefore, they need to develop risk management competencies to identify and mitigate the risks associated with third-party partnerships (e.g., project delays or data security issues). Third, given the lean processes and decision-making structures, SME employees need to establish open lines of communication with external stakeholders themselves (Zavo, 2022). They need to convey technical information to non-technical stakeholders by understanding and resonating with their concerns and needs.

## PRACTICAL RECOMMENDATIONS: A COMPASS FOR GAI DEPLOYMENT IN SMES

SME leaders should view GAI as a revolutionary, general-purpose technology like electricity or the internet (McAfee et al., 2023). They should also recognize the looming challenges of creating significant value through GAI technology, as past studies indicate that most AI investments have little or no impact on business value (Davenport & Ronanki, 2018; Ransbotham et al., 2020). We believe that under the right conditions, GAI can streamline work processes and unleash innovation to result in long-term competitiveness for SMEs. To achieve this, SMEs must respond quickly and resourcefully to technological opportunities and equip themselves with the ability to adapt to this rapidly changing landscape (Rajaram, 2023). Drawing from the four strategic elements of our sailing metaphor, we advocate key principles that serve as a compass for SMEs' seamless GAI deployment in their business operations.

### Competency of Employees

**Design the right ecosystem to upskill the workforce**
The first strategic element to prioritize is employee competency. We suggest that SMEs should develop an ecosystem that facilitates quick upskilling of technical and non-technical employees through in-person and online training courses, seminars, hands-on practice, and mentoring. The learning program can focus on both technical (e.g., database management, programming, statistical analysis, machine learning, data cleaning and preprocessing, data visualization, and data interpretation) and soft skills (e.g., learning orientation, technology savviness and curiosity, and adaptability). For instance, employees need to learn to avoid taking GAI outputs at face value. As GAI models are prone to "hallucinations", employees need to vet the reasonableness of algorithmic outcomes and verify whether they accomplish the desired tasks (McAfee et al., 2023). Furthermore, while academic institutions are expanding their teaching curricula in the ML area, it can take many years for this to have an impact on the workforce and ML-educated employees remain scarce. Therefore, SMEs can propose in-house development programs for existing technical employees, for instance through certification programs in data analytics (Lee & Shin, 2020). While SMEs may not be able to afford the heavy investments for setting up such programs, they could rely on Massive Open Online Courses (MOOCs), Edx, LinkedIn Learning, DataCamp, Google AI, Kaggle, Microsoft Learn and Udacity, or ask industry unions to provide resources to upskill their technical employees.

**Find the expertise beyond the walls of the SME**

Firms should build cross-functional teams that include AI specialists, domain experts, and system users to facilitate GAI deployment (Jarrahi et al., 2023). However, expertise is typically not readily available in SMEs, which implies that cross-functional teams may need to span organizational boundaries. For instance, SMEs may rely on GAI champions from close business partners who function as boundary spanners between GAI capabilities and business needs. They may also participate in GAI open-source communities (Shrestha et al., 2023) to facilitate GAI customization and deployment.

<div align="center">

**Effective Leadership and Work Values**

</div>

**Communicate a clear vision with fortitude**

Leadership is vital in driving GAI deployment. SME leaders need to show how switching to novel forms of collaboration with GAI can enable the SME to reach its growth goals. They must share their vision by demonstrating fortitude and mental and emotional strength of character to overcome setbacks without losing sight of one's values or objectives. They must also explain why GAI is a strategic necessity for the SME and rally employees around a common goal. SME leaders should negotiate and obtain employees' buy-in by fully understanding and addressing their concerns. For instance, GAI's advanced functionalities can partially automate employees' work tasks, which puts them in a very vulnerable position. SME leaders need to mitigate employees' fear of being replaced by GAI and act as empathetic decision makers (Kaplan & Haenlein, 2020). They should identify those employees who successfully leverage GAI and ask them to inspire the workforce through narratives and demonstrations of the power of the technology. SME leaders must also redesign roles and job responsibilities, partnering closely with human resource professionals and the employees themselves. They should involve employees at an early stage to reduce uncertainty, encourage open discussions, and obtain employees' commitment to the change process.

**Empower employees for the GAI deployment**

Empowering employees for GAI deployment involves creating a work environment that encourages experimentation, innovation, autonomy, and feedback. There are three strategies to make this feasible. 1) Support innovation and develop scaffolded structures to encourage exploration and creative thinking. For instance, SME leaders can recognize and reward employees who have shown substantial creativity supported by GAI. They can encourage employees to experiment with GAI solutions to resolve their business challenges and infuse interest by engaging their employees through a fun and light-hearted approach (e.g., GAI hackathons or innovation challenges). 2) Foster employee autonomy. SME leaders should trust their employees to make decisions on how they can efficiently collaborate with GAI within their scope of work. 3) Establish feedback channels. SME leaders should provide channels for employees to offer their honest feedback on GAI tools. SME leaders can then collate and constructively use this feedback to improve GAI processes, structure, systems, and workflows.

**Become powerful change agents**

SME leaders must become visible and powerful change agents, promoting the novel collaboration possibilities of GAI. To increase adoption, they must demonstrate that deploying GAI follows essential guiding principles, such as compliant data management practices, transparent decision-making processes on model architectures and data sources, clear accountability mechanisms, and strong ethical principles. SME leaders must also walk the talk and act as role models in this transformation process. For instance, they can demonstrate how they personally use GAI and share their own experiences with the limitations and challenges of the technology. SME leaders should also demonstrate agility by thinking creatively on how employees may better integrate GAI in their work and experimenting with further scaling opportunities. They can promote agile methods to successfully deploy GAI, as multiple iterations can quickly evaluate hypotheses and develop understanding of the most promising application areas (McAfee et al., 2023). Learning through multiple iterations is required to develop effective interactions with LLMs, a field called "prompt engineering" (Robertson, 2024) that is becoming a vital aspect of responsible, value-added, and constructive GAI deployment. Primarily, it is an expert process that enables users to intentionally shape the behavior of GAI models to the expected outcomes. This comprises the action of iteratively crafting effective and accurate prompts to obtain desired responses from GAI models. It empowers users to customize GAI outputs to explicit tasks and domains, enhance work efficiency, and alleviate potential biases.

## Organizational Culture, Collaboration, and Cooperation

**Nurture an effective and healthy organizational culture**

SMEs must cultivate an organizational culture that embraces technological adoption. Depending on factors such as the focal industry, the goals of the SME or its business development phase, the SME can lean towards a market or adhocratic culture. SMEs can foster a market culture that emphasizes the attainment of specific goals and focuses on outperforming competitors if they evolve in mature industries with clear standards and regulations. Such a culture drives cost control, incremental innovation, and efficiency. By contrast, SMEs can embrace an adhocracy culture if their work is knowledge-based and involves creativity, such as in design, research, or software development. An adhocracy culture is particularly suited to such contexts as its decentralized decision making and flat hierarchy promote problem solving, creativity, and individual autonomy, which are essential for successfully deploying GAI. An effective and healthy organizational culture should emphasize values that serve as a compass for growth and sustainability and SMEs should incorporate relevant and necessary control mechanisms to instill such a culture. Values such as empowerment, accountability, and trust must be rooted in the culture to effectively deal with GAI deployment. For instance, SMEs can infuse values such as empowerment and accountability in decision making through customizing the user interfaces that provide GAI-generated suggestions. SMEs can facilitate work processes that expect employees to make informed decisions by altering the way GAI-generated suggestions are shown to users and suggesting they check alternative sources before making final decisions. This approach advocates a culture of empowerment and holds employees accountable for their own actions. Moreover, customizing the user interface provides users with more control over the GAI system's behavior to better address their needs (Frissora, 2021).

**Address GAI challenges through collaboration and cooperation**

SMEs need to cope with important adoption hurdles while using existing GAI interfaces and/or customizing GAI applications, such as mitigating the threat to employees' professional identities, dealing with the complexity and opacity of algorithms, and reducing algorithmic aversion. They must also identify the strategies that increase the accuracy of predictions and opt for precautionary measures to address reputational, legal, and ethical concerns. Addressing the complex, multifaceted challenges arising in the transition to human-GAI collaboration requires SME leaders and employees to have good conflict management, problem-solving, negotiation and alignment skills. Engaging all stakeholders to mitigate the challenges surrounding GAI deployment is a necessary condition for SMEs' productivity and long-term competitiveness in such an evolving landscape.

## Relationships with Third Parties

**Take or shape? Decide the degree of reliance on third-party providers**

SMEs must design sound business relationships and cooperation models by addressing their degree of reliance on third parties for GAI solutions. SMEs must carefully choose between a "take or shape" approach while deploying GAI. The "take" approach focuses on using off-the-shelf solutions that are convenient and cost-effective. However, those solutions can lack customization, which could provide a competitive edge in the market; offer weak regulatory compliance and data privacy; and create a strong reliance on a single or a few third-party providers. For instance, SMEs active in software development can adopt a "take" approach by benefitting from instantaneous code generated by LLMs (e.g., GitHub's Copilot), which rely on human coding knowledge. By contrast, the "shape" approach consists of customizing GAI to make it perform specific tasks in niche areas with better relevance and accuracy. This approach enables more customized solutions that address complex and multifaceted needs, yet requires deep and integrated domain expert knowledge, ample time, and complementary resources. For instance, SMEs can leverage open-source AI, which enables developing AI products based on existing tools that large companies (e.g., Meta) have shared as open-source software (Heaven, 2023). Such an approach enables the relative democratization of GAI, as SMEs can download, modify, and deploy the code, which grants a lower degree of dependency on third-party providers than the "take" approach. SMEs can also partner with third parties to develop customized GAI tools and access hardware and software infrastructure. Finally, a mid-range solution consists of creating their own GPTs (i.e., custom versions of ChatGPT) that feature elaborated prompts or specific knowledge yet do not require any coding skills. The choice of approach should account for the SME's strategic goals, risk appetite, resource availability, talent pool, and its novel value proposition and intended competitive edge.

## CONCLUSION

The rise of GAI is quickly disrupting the competitive landscape of SMEs, which are entering uncharted, choppy waters. In this essay, we suggest a sailing metaphor that focuses on four strategic dimensions for GAI deployment: competency of employees, effective leadership and work values, organizational culture, collaboration and cooperation, and relationships with third parties. Our framework adds to practice-oriented scholarship on AI management that does not provide recommendations on how firms can deploy the latest LLMs (e.g., Kaplan & Haenlein, 2020; Leyer & Schneider, 2021; Paschen, Pitt, & Kietzmann, 2020) nor addresses the specific challenges of SMEs in AI deployment (e.g. Bughin, 2023, Akter et al, 2021). We extend this literature by pointing to the important competencies of employees and leaders, as well as the effective collaboration and cooperation required for GAI implementation. Broad AI adoption not only relies on AI technology and skilled employees, but requires that SMEs align their culture, structure, and ways of working (Fountaine et al., 2019). Our work recognizes the socio-technological character of AI, for instance by claiming that GAI can profoundly reshape employees' professional identities. Therefore, we extend prior arguments on the unintended consequences of AI, such as transforming expertise in organizations and reshaping the boundaries of occupations (Holmström, 2022).

Furthermore, Lee and Shin (2020) emphasized the trade-off between accuracy and interpretability of ML algorithms and discussed how organizations can select the appropriate ML algorithm for specific tasks depending on the data type, desired interpretability, and accuracy. Canhoto and Clear (2020) adopted a risk perspective to explain which components of an AI solution can hinder value creation or even lead to value destruction. Our study adds to this scholarship, which has focused on the technical risks pertaining to input data, processing algorithms, and output decisions, by articulating the social risks of deploying GAI. Specifically, we argue that SME employees and leaders can reduce social risks and facilitate the success of GAI through their competencies (see Table 3).

We claim that firms' technical resources are less of a concern for GAI than for past AI developments, because they can be easily accessed via the cloud, software-as-a-service, and application programming interfaces and apps (McAfee et al., 2023). Therefore, we add to Desouza and colleagues (2020) who argued that organizations with immature technical capabilities can benefit from AI in automating routine tasks and improving efficiency. For instance, in billing workflows, SMEs with low technical capabilities can leverage GAI to summarize financial statements and customize recurring bills for individual clients. Similarly, in the inventory management workflow, GAI can create personalized forecasts of stock levels and reordering products lists. GAI can also generate customized reports for product expiration, overall inventory trends, forecasts of product needs and human resources to improve the accuracy of tactical managers' decisions.

We claim that SMEs with low technological maturity can leverage off-the-shelf GAI solutions or partner with third-party technology providers to benefit from GAI in tackling even bolder issues, such as improving scalability and innovation, which may lead to a long-term competitive advantage.

In conclusion, SMEs need to understand and embrace the latest developments in GAI to successfully navigate this tumultuous landscape. Our paper offers a helpful compass for resource-constrained SMEs deploying GAI. Successfully doing so can help them narrow the knowledge and technological gaps they have with the largest firms. It may enhance scalability, innovation, and long-term competitiveness, enabling SMEs to stay afloat and breeze through those uncharted, choppy waters.

# TABLES AND FIGURES

## Table 1: Promises of GAI for SMEs

| GAI Promises for SMEs | Description | Practical Examples |
|---|---|---|
| **Transforming Roles and Boosting Productivity** | GAI is poised to transform roles and boost productivity across functions such as sales and marketing, customer operations, and software development, hence adding trillions of dollars in value to the global economy (McKinsey, 2023). GAI can augment both tactical (functional) and operational tasks. | SMEs can adopt GAI to enhance the administrative workflow and processes. For instance, generative writing tools like ChatGPT can automatically create original market reports with little human intervention. Another example is a small Swiss chocolate confectioner that uses GAI to customize its advertising material and translate it into many languages to scale up its business globally. |
| **Accessing Human Knowledge** | GAI draws from large training datasets that encode significant portions of human knowledge. Hence, GAI tools can generate a wide array of credible writing in seconds and make the writing more fit for purpose based on users' feedback. | SMEs that are active in software development can benefit from the code generated by LLMs (e.g., GitHub's Copilot), which relies on human coding knowledge. |
| **Facilitating Creativity and Empowering Innovation** | GAI is an important force in democratizing creativity (AI TechPark, 2023). GAI can expand the repertoire of possible outputs and provide insights that humans would not have considered. As innovation is the lifeblood of SMEs, leveraging GAI in creative tasks can benefit SME competitiveness. | Creatives (e.g., marketers, content creators) in SMEs can improve the creative process by leveraging GAI to generate ideas based on explicit concepts, color palettes, and draft sketches. For instance, tools such as Midjourney, DALL-E, or Stable Diffusion can generate images from text descriptions (prompts) and augment their work by generating compelling visuals for a marketing campaign. These GAI tools enable creatives to save time, effort, and energy and focus on expressing their ideas. |
| **Increasing Differentiation of Products and Services** | As they typically operate in smaller market segments than their larger counterparts, SMEs can use GAI to differentiate their products and services and better address their markets. | SMEs in the fragrance and beauty care industry can use GAI tools to explore new raw material combinations and customize fragrances and beauty care products. For instance, GAI can facilitate the exploration of product possibilities for specific target market segments, which augments differentiation and creates more value for customers. |
| **Accelerating Product Development** | GAI can accelerate product development by enabling faster iterations and experimentation, which increases market responsiveness. | GAI tools can assist in optimizing and speeding up various aspects of product design and development, thereby reducing costs and improving operational efficiency. For instance, resource-constrained SMEs can use GAI to quickly generate multiple versions of product design prototypes and experiment with the most promising ones. |
| **Supporting Decision Making** | GAI empowers SMEs' higher accuracy in decision making (i.e., based on data evidence, validated information). It can support decision making in business strategy planning and formulation, business operations, growth, and performance sustainability, which reduces the knowledge gap with larger competitors. | SMEs can utilize GAI that facilitates data-driven, agile, and proactive decision making for immediate business impact. For instance, SMEs in precision medicine can leverage GAI to customize models and treatment plans using individual physiological data and family medical history that are available in electronic forms (Kuzlu, Xiao, Catak, Gurler & Guler, 2023). |

**Table 2: Challenges of GAI for SMEs**

| GAI Challenges for SMEs | | Description | Practical Examples |
|---|---|---|---|
| Adoption Hurdles | Threat to Professional Identities | Professional identities impact the way employees accept or ignore AI outputs (Lebovitz et al., 2022). Whenever employees identify strongly with their profession, they will protect their autonomy and strive to enhance their status. Employees may refrain from using GAI if it prevents them from exhibiting expertise in a certain field. | Despite the benefits that GAI can provide for creative tasks, marketing professionals may be reluctant to use GAI to design their advertising visuals. They may strongly identify with such tasks, which give them visibility and status within the SME, and they may feel threatened if GAI can perform them. |
| | Complexity and Opacity | To mimic human-like cognitive processes, such as conflict resolution, problem solving, and reasoning, GAI relies on a large number of parameters and multiple computational steps, which makes algorithms inherently complex. The complexity and opacity of ML algorithms powering GAI is one of the challenges for accelerated adoption (Chui et al., 2018). | GAI outputs may not be explainable by assessing individual parameters or training data. If SME employees cannot understand the way algorithms reach specific outputs, they may resist GAI adoption. |
| | Algorithmic Aversion | Employees' preference for human-made predictions despite the obvious superiority of predictions generated by an algorithm. | Employees tend to decrease their reliance on AI systems when they observe them perform and inevitably err. In performance prediction tasks, evaluators tend to abandon AI systems in favor of a human judge when they realize that such systems are imperfect, even though they outperform human judgment (Dietvorst et al., 2015). |
| Accuracy of Predictions | Random Inaccuracies or Falsehoods | GAI models are prone to "hallucinations," where their outputs contain "random inaccuracies or falsehoods expressed with unjustifiable confidence" (Mukherjee & Chang, 2023; Hannigan et al., 2024). | A New York attorney from a small law firm relied on a GAI chatbot to extract case law citations for a personal injury case against an airline (Weiser, 2023). The chatbot provided legal cases that were fabricated and could not be found in reputable legal databases. The judge eventually fined the attorney for failing to verify his sources. |
| | Learning from Existing Biases | Algorithmic outputs can perpetuate and even amplify organizational biases in important practices of SMEs, resulting from unwanted biases in the training data (Cowgill, 2019). Therefore, SMEs need to mitigate such biases and reduce social inequalities and discrimination in organizational processes. | GAI chatbots offering career counselling in human resource management can generate discriminatory or biased advice. For instance, the chatbots can fail to recommend relevant programming courses to certain demographics (e.g., women), because the training dataset may be biased against such groups (due to a historical small number of women taking such courses). |
| Ethical Issues | Accountability for Errors or Misjudgment | Accountability is murkier when it comes to the consequences of errors or misjudgment, as GAI systems, developers, and employees are all involved in the decision-making process. | Managers in SMEs can rely on GAI for their business-related decision making. For instance, finance managers can use GAI-automated reports that advise on investments in financial markets. Algorithmic recommendations might lead to ill-judged investment decisions, the accountability for which remains unclear. |
| | Trust and Authenticity Concerns | GAI blurs the line between human-generated and computer-generated content, which causes trust and authenticity concerns. | For instance, customers may be unaware of whether they are communicating with a conversational GAI chatbot or a human (e.g., textually or orally). As GAI interactions can appeal to emotions from users, GAI could exploit vulnerabilities of human psychology in unethical ways. |
| **Reputational and Legal Risks** | **Misuse of Proprietary Data** | Training data can be protected by intellectual property laws, which creates a rather murky legal landscape whenever GAI that has been trained on proprietary data generates new content. | SMEs can leverage GAI-powered image generators to increase creativity in the design of marketing visuals. Yet, many tool providers have trained their models on proprietary images scraped from the internet without asking for the consent of the original artists. While the legal consequences of sharing content generated by such image generators are still murky, some image generators use training data that are copyright-free, open licensed, or owned in-house to avoid any legal issues (e.g., Adobe Firefly). |
| | **Vulnerability of Confidential and Sensitive Data** | SMEs need to maintain and update cyber protection systems against criminal acts (e.g., stealing confidential and sensitive data). Failing to do so may potentially lead to reputational damage or the possibility of being entangled in legal risks. | There is a possibility that data from millions of users used to fine-tune LLMs can be stolen and utilized for bad purposes. For instance, data of an SME-sized medical service provider offering its services to 100,000 people can fall into the hands of cyber scammers. The data is extremely sensitive, as it can potentially include patients' personal particulars, medical history, health problems, and much more. |
| **Reliance on Third Parties** | **Lock-in Effect** | Due to limited expertise, SMEs can outsource GAI solutions to third-party GAI solution providers. | These GAI solution providers may change their business conditions (e.g., pricing models) or discontinue their services without consulting SMEs. This lock-in situation may prevent SMEs from optimizing their business operations and securing market opportunities. |
| | **Inadequate Services** | GAI solution providers may not to be accustomed to SMEs' niche business areas, which require adequate knowledge to design appropriate GAI solutions effectively. | GAI solution providers may not have the relevant resources to provide speedy and comprehensive services and support. For instance, SMEs using a GAI-powered translation service may experience delays when interpreting a spoken language in real-time, making it impractical for live conversations. During peak usage times, the service may struggle to deal with the volume of translation requests, leading to slow performance and increased latency. |
| **Financial Burden** | **Investment in Process Infrastructure** | SMEs need to invest in the relevant process infrastructure to incorporate GAI into the operational ecosystem and achieve employee re-skilling. Employees need to be upskilled to embrace the newly minted GAI-enhanced work environment. | SMEs can leverage on economical online courses (e.g. MOOCs) to upskill and re-skill employees. They can also use Edx, LinkedIn Learning, DataCamp, Google AI, Kaggle, Microsoft Learn and Udacity, or ask industry unions to provide resources. SMEs can develop an ecosystem that facilitates quick upskilling of technical and non-technical employees through in-person and online training courses, seminars, hands-on practice, and mentoring. |

**Table 3: Strategic Elements of the Sailing Metaphor for GAI Deployment in SMEs**

| Strategic Elements for GAI Deployment | Sailing Metaphor | SME Stakeholders | SME Specificities | Competencies Required for GAI Deployment | Description |
|---|---|---|---|---|---|
| **Competency of Employees** | **Quality of the crew** | Employees | SME employees need to learn on the fly with few structured training programs. They need to find solutions and learn by themselves (Haryanto, Hatyono & Sawitri, 2017). | **Learning orientation** | Ability to continuously unlearn and learn to work in the novel human-GAI collaboration system. |
| | | | SME employees have hands-on experience with GAI deployment and some degree of decision-making authority over GAI use. (Gonzalez-Varona, Acebes, Poza & Lopez-Paredes, 2020) | **Technology savviness and curiosity** | Ability and interest to experiment with novel technological features and stay up to date with rapidly evolving technological advancement. |
| | | | SME employees wear multiple hats and need to handle multiple aspects of GAI deployment (Wishart, 2018) | **Adaptability** | Ability to realign and meet the rapidly evolving needs arising from GAI deployment. Positive attitude despite the discomfort to understand new work demands and adjust behavior for the collective enhancement of the SME. |
| **Effective Leadership and Work Values** | **Quality of the ship captains** | Leaders | SME leaders need to demonstrate fortitude to endure, deal with and overcome obstacles or hardships in GAI deployment (Matinez-dei-Rio, Perex-Luno & Bojica, 2023). | **Fortitude** | Courage, mental and emotional strength to face challenges with perseverance, steadfastness, and a strong determination to deal with adversity in GAI deployment (e.g., business disruptions, financial burden, employee resistance). |
| | | | SMEs can embrace technological change, yet opportunities for value creation are constrained by tight resources. SME leaders need to make strategic decisions on resource allocation, accounting for cost and potential (Walt, Kroon & Fourie, 2015). | **Vision** | Foresight to appreciate and leverage the opportunity of GAI's powerful capabilities for scalability, creativity, and business growth within tight resource constraints. |
| | | | SME leaders oversee multiple functions of the SME. They need to ensure seamless integration across different departments and alignment with the broader business goals (Menkhoff, Loh & Kay, 2023). | **Change agent** | Becoming an ambassador for buy-in and empowering employees to appreciate the benefits of GAI for creativity and scalability. Encouraging an innovative mindset to view and embrace GAI as beneficial to all stakeholders. For example, making employees realize how their current role can be enhanced with GAI. |

| | | | SME leaders can quickly respond to evolving market conditions given the flexible operational processes and structures of the SME (Walter, 2021). | **Agility** | Quickly addressing GAI's disruptions in the workflow with focused and high-quality solutions. |
|---|---|---|---|---|---|
| **Organizational Culture, Collaboration and Cooperation** | **Navigation of the ship** | **Employees and leaders** | As the organizational structure of SMEs is rather lean, SME employees and leaders need to collaborate closely and effectively in the GAI deployment (Mohiuddin, 2017; Karadakal, Goud & Thomas, 2015). | **Conflict management and problem solving** | Ability to deal with and address issues arising due to GAI disruptions with tact yet objectivity. |
| | | | As teams are typically small and hierarchies flat, cross-functional collaboration is essential to gather relevant insights and align GAI initiatives with business goals (Major, 2023). | **Negotiation and alignment** | Ability to balance different perspectives on GAI. Ability to find common ground and help everyone to see the holistic benefit for the SME. |
| **Relationships with Third Parties** | **The sailing route** | **Employees and external stakeholders** | SMEs typically lack established processes and infrastructure that support collaboration with external parties (Meister, 2006). | **Project management** | Ability to define goals and milestones and monitor timelines to make sure that GAI deployment stays on track. |
| | | | Due to their entrepreneurial mindset and minimal bureaucracy, many SMEs have limited resources and expertise for risk management, contracting and compliance for developing novel technological solutions with third parties (Bluebox Content Team, 2024). | **Risk management** | Ability to identify and mitigate risks associated with third-party partnerships, such as project delays and data and cyber security issues. |
| | | | Given the lean processes and decision-making structures, SME employees need to establish open lines of communication with external stakeholders themselves (Zavo, 2022). | **Communication** | Ability to convey technical information to non-technical stakeholders by understanding and resonating with their concerns and needs. |

**Figure 1: The Sailing Metaphor – SMEs Navigating Uncharted, Choppy Waters**
*Generated by Adobe Firefly with the prompt "A small boat navigating agitated waters in a heavy storm"*

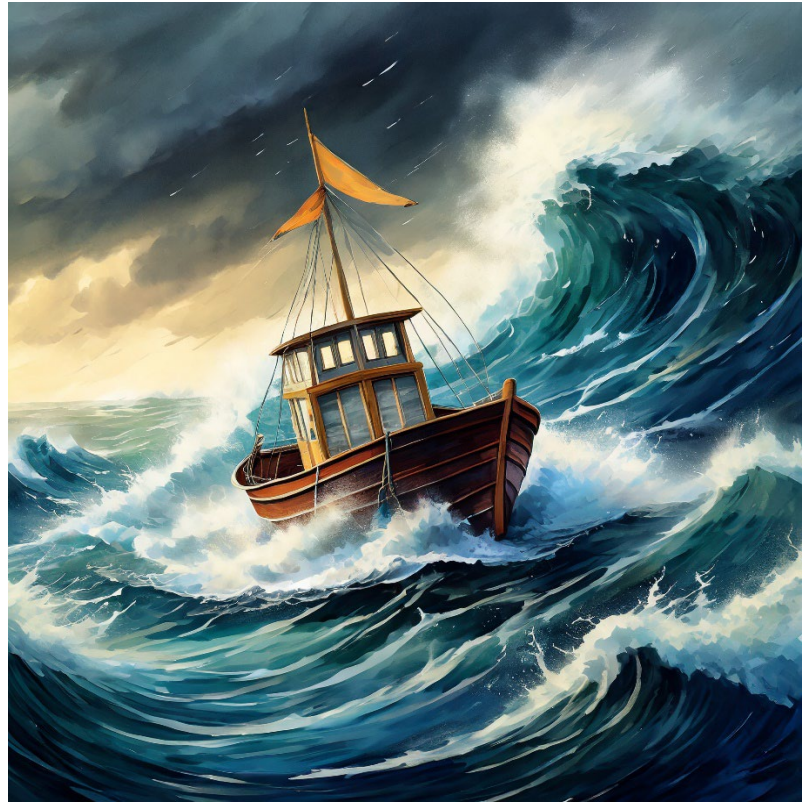